\documentclass[12pt]{article}

\usepackage{tikz, adjustbox, float} % For diagrams
\usepackage[round]{natbib} % For references
\usepackage{amsmath, amsfonts, amssymb, amsthm, dsfont} % For math symbols
\usepackage{mathrsfs} % For additional caligraphic letters
\usepackage{mathabx} % For \widecheck
\usepackage{soul} % For crossed text
\usepackage{bm} % For bold math
\usepackage{booktabs}  % For better tables
\usepackage{enumitem} % For enumerate alphabet
\usepackage{algorithm,algorithmic}
\usepackage{mathtools}

\def\0{{\bf 0}}
\def\1{{\bf 1}}

\def\x{{\bf x}}

\DeclareMathOperator*{\var}{var}

\DeclareMathOperator*{\argzero}{argzero}

\usepackage{latexsym}
\usepackage{amsthm}
\usepackage{amsmath,amssymb,amsfonts,bm,amsbsy,bbm}
\usepackage{epsfig}
\usepackage{graphicx,rotating,lscape}
\usepackage{hyperref}
\usepackage{verbatim}
\usepackage{natbib}
\usepackage{multirow,color}
\usepackage{geometry}
\usepackage{setspace}
\usepackage{subfigure}
\usepackage{bbm}
\usepackage{ulem} % For strickout in math mode

\def\x{{\bf x}}

\def\bb{{\boldsymbol\beta}}

\def\0{{\bf 0}}

\def\var{\hbox{var}}

\def\bse{\begin{eqnarray*}}
\def\ese{\end{eqnarray*}}
\def\be{\begin{eqnarray}}
\def\ee{\end{eqnarray}}
\def\bsq{\begin{equation*}}
\def\esq{\end{equation*}}
\def\bq{\begin{equation}}
\def\eq{\end{equation}}

\def\boxit#1{\vbox{\hrule\hbox{\vrule\kern6pt  \vbox{\kern6pt#1\kern6pt}\kern6pt\vrule}\hrule}}
\def\bse{\begin{eqnarray*}}
\def\ese{\end{eqnarray*}}
\def\be{\begin{eqnarray}}
\def\ee{\end{eqnarray}}
\def\bsq{\begin{equation*}}
\def\esq{\end{equation*}}
\def\bq{\begin{equation}}
\def\eq{\end{equation}}

\def\var{\hbox{var}}

\def\x{{\bf x}}

\def\squarebox#1{\hbox to #1{\hfill\vbox to #1{\vfill}}}

\def\0{{\bf 0}}

\def\var{\hbox{var}}

\newtheoremstyle{mytheoremstyle} % name
    {0.3cm}                      % Space above
    {0cm}                        % Space below
    {}%{\it}                   % Body font
    {}                           % Indent amount
    {\bf}                   % Theorem head font
    {: }                          % Punctuation after theorem head
    {0em}                       % Space after theorem head
    {}  % Theorem head spec (can be left empty, meaning normal)

\theoremstyle{mytheoremstyle}

\newtheorem*{Lemma*}{Lemma}

\newtheoremstyle{myExampleRemarkstyle} % name
    {0.3cm}                    % Space above
    {0cm}                      % Space below
    {\it}                         % Body font
    {}                         % Indent amount
    {\bf}                      % Theorem head font
    {: }                       % Punctuation after theorem head
    {0em}                      % Space after theorem head
    {}  % Theorem head spec (can be left empty, meaning normal)

\theoremstyle{myExampleRemarkstyle}
\newtheorem{Example}{Example}

\newtheorem{innercustomgeneric}{\customgenericname}
\providecommand{\customgenericname}{}
\newcommand{\newcustomtheorem}[2]{%
  \newenvironment{#1}[1]
  {%
   \renewcommand\customgenericname{#2}%
   \renewcommand\theinnercustomgeneric{##1}%
   \innercustomgeneric
  }
  {\endinnercustomgeneric}
}

\newcustomtheorem{customthm}{Theorem}
\newcustomtheorem{customlemma}{Lemma}
\usepackage{mathalfa}
\usepackage{oubraces}

\let\refBKP\ref
\renewcommand{\ref}[1]{{\upshape\refBKP{#1}}}

\usepackage{setspace}
\definecolor{green}{rgb}{0.15,0.6,0.3}
\hypersetup{
citebordercolor=white,
filebordercolor=red,
linkbordercolor=green
}
\hypersetup{
colorlinks,
citecolor=green,
linkcolor=red,
urlcolor=green}

\begin{document}

\begingroup  
\begin{center}
%\Large Exploiting multi-design outcomes to increase replicability: Introducing the g-value 
\LARGE Global p-Values in Multi-Design Studies 
%\LARGE Defining Global p-Values in Multi-Design Studies 
%\LARGE From Forking Paths to Replicability: \\ A Step Toward Better Multi-Design Studies
\vspace{1cm}\\
\large{Guillaume Coqueret$^{1}$, Yuming Zhang$^{2}$, Christophe Pérignon$^{3}$, \\ Francesca Chiaromonte$^{4,5}$, Stéphane Guerrier$^{6}$}
\vspace{0.6cm}\\
\small{$^{1}$Department of Quantitative Finance and Economics, EMLYON Business School, France \\ 
$^{2}$Department of Biostatistics, Harvard University, USA \\ 
$^3$Department of Finance, HEC Paris, France \\
$^{4}$Institute of Economics and L'EMbeDS, Scuola Superiore Sant'Anna, Italy \\ 
$^{5}$Department of Statistics, Pennsylvania State University, USA \\ 
$^{6}$Faculty of Science, University of Geneva, Switzerland}
\end{center}
\endgroup

\vspace{-0.2cm}  
\begin{abstract}
Replicability issues -- referring to the difficulty or failure of independent researchers to corroborate the results of published studies -- have hindered the meaningful progression of science and eroded public trust in scientific findings. In response to the replicability crisis, one approach is the use of multi-design studies, which incorporate multiple analysis strategies to address a single research question. However, there remains a lack of methods for effectively combining outcomes in multi-design studies. In this paper, we propose a unified framework based on the \textit{g-value}, for global p-value, which enables meaningful aggregation of outcomes from all the considered analysis strategies in multi-design studies. Our framework mitigates the risk of selective reporting while rigorously controlling type I error rates. At the same time, it maintains statistical power and reduces the likelihood of overlooking true positive effects. Importantly, our method is flexible and broadly applicable across various scientific domains and outcome results. %We showcase its practical utility through a real-world application in pharmaceutical science.
\vspace{0.8cm}

\end{abstract}
\newpage

\section{Introduction}

Trust in results forms the bedrock of scientific progress; without it, science cannot advance in a cumulative or meaningful way. Yet across many fields today, empirical findings are often disputed or overturned by subsequent studies. Independent attempts to reproduce published results frequently fail, even when researchers employ established methodologies on new data. This persistent pattern of inconsistency -- commonly described as the replicability crisis -- undermines the credibility of scientific knowledge in the eyes of society. This crisis spans a diverse array of disciplines, including medicine \citep{ioannidis2005most}, psychology \citep{shrout2018psychology}, political science \citep{king1995replication}, social sciences \citep{Camerer_al18socialscience}, and economics \citep{Camerer_al16}.

%Several causes for the lack of external validity of some published results have been identified. One is the (often unintentional) use of incorrect or inappropriate statistical analyses, which can lead researchers to draw unreliable conclusions from their data. The proliferation of complex data and analytic tools has, in some ways, exacerbated this problem, as it becomes increasingly easy to generate results without a full appreciation of the underlying limitations and potential sources of bias. When such statistical errors go unnoticed, they can propagate misleading conclusions throughout the literature and hinder scientific progress.

A major contributor to the replicability crisis is the persistence of questionable research practices, particularly selective reporting and publication bias. Indeed, researchers often face pressure to produce statistically significant results to increase their chances of publishing in high-impact academic journals. This pressure can incentivize behaviors such as testing multiple model specifications or data treatments until a desirable p-value is obtained -- a practice known as p-hacking. Empirical evidence for p-hacking includes the under-representation of marginally insignificant results in the published literature \citep{brodeur2016star, brodeur2020methods}, as well as patterns revealed through p-curve analysis \citep{simonsohn2014p, elliott2022detecting}. Selective reporting -- where only results supporting a particular hypothesis are disclosed, while null or non-significant findings are omitted -- is also known as the file-drawer problem. This bias is further exacerbated by editorial preferences for novel or positive outcomes and has been widely documented across scientific fields \citep{easterbrook1991publication, stanley2005beyond, john2012measuring, banks2016evidence, xie2021prevalence}.

Several methods have been proposed to mitigate the replicability crisis.  For instance, preregistration by researchers of their hypotheses, methods, and analysis plans before collecting data is a valuable tool \citep{miguel2014science, simmons2021pre}. It helps mitigate certain questionable research practices, such as p-hacking, selective reporting, and HARKing (Hypothesizing After the Results are Known). However, preregistration alone cannot (i) ensure quality research design as poorly designed studies can still be preregistered, (ii) guarantee external validity of the reported results, and (iii) address analytical flexibility entirely as researchers can preregister vague plans or exploit loopholes.\footnote{Building on \citet{simonsohn2014p}, consider a researcher who preregisters a target sample size of 200 participants but engages in p-hacking by repeatedly analyzing the data after every five additional participants are added, stopping data collection as soon as statistical significance is achieved and ultimately reporting results based on the most recent 200 participants.}

Another particularly promising approach involves conducting \textit{multi-design studies}. In this approach, different researchers independently examine the same research question using a variety of analysis strategies (\textit{multi-analyst studies}) or a single researcher examines a given research question using a variety of reasonable analysis strategies (\textit{multi-path studies}). Such analysis decisions can vary the data collection procedure, selection and transformation of the variables, modeling method, and so on. By pooling insights from different analysts or protocols -- similar to a ``wisdom of the crowds'' effect -- such studies can internalize the uncertainty arising from the various methodological paths. This approach also helps reduce the risk of p-hacking, generate a more thorough assessment of the robustness of findings, and increase confidence in the research conclusions.\footnote{The impact of protocol variation has been extensively documented in recent empirical research across disciplines, including psychology \citep{silberzahn2018many}, biology \citep{gould2025same}, neuroimaging \citep{botviniknezer2020}, economics \citep{huntington2021influence}, and finance \citep{menkveld2024nonstandard}.} 

However, it remains unclear how to rigorously exploit the myriad of outputs generated from multi-design studies. While newer methods, such as p-value aggregation, can preserve type~I error rates, they often suffer from limited statistical power, rely on arbitrary functions, or may lack theoretical guarantees \citep{wilson2019harmonic, vovk2020combining, yoon2021powerful}. \cite{simonsohn2020specification} explores how to use all outcomes from different analytic specifications to evaluate their consistency with the null hypothesis of no effect. They propose three test statistics for conducting specification curve analysis. Our approach differs in two key ways: (i) we operate directly on p-values rather than effect sizes, and (ii) our method for aggregating p-values does not rely on a continuous transformation as in Stouffer's method. Alternatives like stricter confidence levels or Bayesian adjustments \citep{Johnson2013standards,benjamin2018redefine,mccloskey2025critical} can guard against false positives but risk increasing false negatives. Therefore, while the existing literature has advanced in identifying the problem and proposing potential remedies to account for the multiple analysis decisions, a comprehensive and robust solution that unifies these approaches remains to be established.

In this paper, we propose a unified framework for integrating a variety of analytical outcomes, particularly the p-values, to support inference and decision-making. In particular, we introduce the \textit{g-value}, for global p-value. This novel metric extends the traditional p-value for use in multi-design studies, enabling meaningful combination of results from diverse analysis strategies. Our framework is able to mitigate the risks associated with selective reporting while rigorously controlling the type I error rates. Meanwhile, it preserves statistical power and reduces the possibility of missing true positive effects. %Importantly, our method is flexible and broadly applicable across various domains, including economics, finance, and the life and medical sciences. We showcase its practical utility through a real-world application in pharmaceutical science.

It is important to emphasize that our approach differs fundamentally from the methods typically discussed in the multiple testing literature. That body of work (see e.g., \citealt{benjamini1995controlling,gasparin2025combining}) typically addresses the challenge of testing numerous hypotheses across different variables, as commonly encountered in fields like genomics. Our work is also distinct from the selective inference literature (see e.g., \citealt{taylor2015statistical}), which seeks to adjust for the reuse of data by a single analyst for both model selection and inference. Notably, the literatures on multiple testing and selective inference generally consider a single analyst making choices over multiple hypotheses or selection events. In contrast, our framework is motivated by the diversity of analysis decisions (e.g., data preprocessing, outcome selection, model specification) that occur when a single research question is examined from multiple analytical perspectives. By synthesizing results across these various analytical paths, our approach advances the principles of open science \citep{Nosek2015, christensen2019transparent} and veridical data science \citep{yu2020veridical}: it encourages transparency, compels researchers to document alternative analysis strategies, and guards against cherry-picking results.

The rest of the paper is organized as follows. Section~\ref{sec:problem} formulates the problem under consideration and highlights the risk associated with naively combining p-values from different analysis strategies. In Section~\ref{sec:method}, we introduce the proposed g-value and demonstrate its application through a simple example. This example illustrates that our method can combine p-values from different analyses in a simple manner, while controlling the type I error rate and preserving strong statistical power. Section~\ref{sec:practical} discusses practical considerations for our proposed framework, including the range of analyses relevant in multi-design settings and practical computation of the g-value. Finally, Section~\ref{sec:conclusion} concludes the paper.

\section{Problem Formulation}
\label{sec:problem}

In this paper, we focus on {\em confirmatory analyses}. In contrast to exploratory analyses, where one might use data to help shape or refine hypotheses, confirmatory analyses use data exclusively to validate a predefined hypothesis. These analyses are widely used across various disciplines. For instance, in the medical field, a phase III confirmatory trial is designed to establish the positive effect of an intervention (e.g., drug, treatment) compared to a reference group in altering a disease state (see e.g., \citealt{boateng2019review}). Similarly, in the social sciences, many studies focus on assessing the influence of a specific individual trait (e.g., race, gender, level of education) on a behavior (e.g., alcohol consumption, drug abuse, violence) (see e.g., \citealt{muller2025race}). % or decision-making processes formulate hypotheses about the expected direction of these effects (see e.g., \citealt{kramer2020race}).
In the context of confirmatory analysis, the predefined hypotheses typically take the form of one-sided tests. Thus, without loss of generality, we consider: 
\begin{equation}
\label{eq:hypotheses}
    H_0:\; \mu = 0 \quad \text{versus} \quad H_1:\; \mu > 0,
\end{equation}
where $\mu$ is the parameter of interest (e.g., the parameter associated to the treatment effect). This formulation covers means, but also, more generally, regression settings -- for example, when studying the effect of a single variable while adjusting for other relevant covariates. Applicable models include generalized linear and mixed models, which are commonly used in medical research, as well as panel data models, frequently employed in economics \citep{matyas2013econometrics}.

We assume that an estimator of $\mu$, denoted by $\hat{\mu}$, can be computed from a sample of size $n$ and that it satisfies:  
\begin{equation*}
\label{eq:asymtptotic:distribution:beta} 
    T_n \vcentcolon= \frac{\sqrt{n} (\hat{\mu} - \mu)}{\hat{\sigma}} \overset{d}{\to}  \mathcal{N}(0, 1) \, ,
\end{equation*}
where $\hat{\sigma}$ is a suitable estimator of the asymptotic standard deviation of $\hat{\mu}$, and $\overset{d}{\to}$ denotes convergence in distribution. This is guaranteed in a broad range of scenarios by relevant central limit theorems.

This problem formulation allows us to define the p-variable $P_n$, based on the behavior of test statistic $T_n$ under the null hypothesis:
\begin{equation*}
\label{eq:pvalue}
    P_n \vcentcolon= \Pr \left( Z \geq T_n \big| H_0, T_n \right), % = 1 - \Phi\left( \sqrt{n} \hat{\mu} / \hat{\sigma} \right),
\end{equation*}
with $Z \sim \mathcal{N}(0,1)$. The realization of the p-variable is called the p-value. Throughout, we use the terms ``p-variable'' and ``p-value'' interchangeably, as well as ``g-variate'' and ``g-value'', which will be defined later. This should cause no confusion in our context. Given a significance level $\alpha \in (0,1)$, the decision rule of the testing procedure is to not reject $H_0$ if $P_n > \alpha$, and to reject $H_0$ if $P_n \leq \alpha$. This approach adequately controls the type I error rate since the asymptotic probability of rejecting $H_0$ given that $H_0$ is correct is equal to $\alpha$:
\begin{equation*}
    \lim_{n \to \infty} \Pr \left( P_n \leq \alpha \big| H_0 \right) = \lim_{n \to \infty} \Pr \left\{  \sqrt{n} \hat{\mu} / \hat{\sigma}  \geq \Phi^{-1} (1 - \alpha) \big| H_0 \right\} = \alpha \, ,
\end{equation*}
where $\Phi(\cdot)$ denotes the cumulative distribution function of the standard Normal distribution.

As previously discussed, the formulation of the hypothesis and the computation of the p-value depend on a number of decisions made by the analyst. Even for a seemingly simple task, such as testing the population mean based on a given sample, the computation of the p-value can be influenced by the specific testing procedure selected by the analyst, which in turn depends on the model assumptions one is willing to make. For example, should one use a t-test or a Wilcoxon rank-sum test? 

Henceforth, we use $A \in \mathcal{A}$ to denote the analyst's choice of strategy. For clarity, we index the corresponding p-value by $A$ and denote it as $P_n(A)$. $\mathcal{A}$ represents the set of (multiple) analysis choices that can be employed to tackle our (single) research question; we denote their total number as $m \vcentcolon= |\mathcal{A}| > 1$ and we assume $m$ to be bounded. Moreover, we assume that the analyst is reasonable and only considers choices that are asymptotically valid and powerful. This means that, for any $A \in \mathcal{A}$, the corresponding test (i) has an asymptotic type~I error bounded by the nominal level $\alpha$, and (ii) has an asymptotic type II error of zero:
\begin{equation*}
\label{eq:asymptotic:valid}
    \lim_{n \to \infty} \Pr\left\{P_n(A) \leq \alpha \big| H_0, A \right\} \leq \alpha \quad \text{and} \quad \lim_{n \to \infty} \Pr\left\{P_n(A) > \alpha \big| H_1, A \right\} = 0 \, .
\end{equation*}
Using these notations, we can see that if the analyst's choice is made \textit{independently} of the data, then we have: 
\begin{equation*}
\begin{aligned}
    \lim_{n \to \infty} \Pr\left\{P_n(A) \leq \alpha \big| H_0\right\} =   \sum_{A \in \mathcal{A}} \Pr(A) \lim_{n \to \infty} \Pr\left\{P_n(A) \leq \alpha \big| H_0, A\right\} \leq \alpha \sum_{A \in \mathcal{A}} \Pr(A) = \alpha.
\end{aligned}
\end{equation*}
However, if the analyst's choice is made \textit{conditionally} on the observed data -- for example, by exploring multiple strategies until a significant result appears -- the risk of false positives can increase. To quantify such risk, we define the \textit{maximum asymptotic size} of $\mathcal{A}$ as: 
\begin{equation*}
\label{eq:max:size}
    s(\alpha, \mathcal{A}) \vcentcolon=  \lim_{n \to \infty} \; \Pr\left\{ \min_{A \in \mathcal{A}} \; P_n(A) \leq \alpha  \big| H_0\right\} \, .
\end{equation*}
Intuitively, this measures the asymptotic type I error rate under an opportunistic scenario, where an adversarial analyst selects the $A \in \mathcal{A}$ affording the smallest p-value, to be compared with the nominal level $\alpha$. Consequently, the asymptotic size of any procedure where $A$ is chosen conditionally on the data is necessarily bounded above by $s(\alpha, \mathcal{A})$.
\begin{Example}
\label{ex:t-test}
Consider a sample of $n$ i.i.d.~observations $X_i \sim F$, with $\mathbb{E}(X_i)=\mu$ and $\var(X_i) = \sigma^2$ both bounded, and a density symmetric around the mean $\mu$.
%size $n$ with $X_i\overset{iid}{\sim} F$, where the probability density function $X_i$ is symmetric around $\mu$. Moreover, we have $\mathbb{E}(X_i)=\mu$ and $\var(X_i) = \sigma^2$, both of which are bounded. 
Based on this sample, we consider the one-sided hypothesis formulated in \eqref{eq:hypotheses} and $m=2$ testing procedures; namely, the one-sample t-test ($A=1$), and the Wilcoxon rank-sum test ($A=2$). By studying the asymptotic joint distribution of the two corresponding test statistics, we can demonstrate that:
\begin{equation*} 
    s(\alpha, \mathcal{A}) = 2\int_{z_{1 - \alpha}}^\infty \varphi(y) \Phi\left\{z(y)\right\} dy 
    > \int_{z_{1 - \alpha}}^\infty \varphi(y) dy = \alpha \, ,
\end{equation*}
where $z(y) \vcentcolon= y (1 - \sqrt{3/\pi}) (1 - {3/\pi})^{-1/2}$ and $\varphi(\cdot)$ is the probability density function of the standard Normal distribution. Therefore, if we fix $\alpha = 0.05$, we obtain $s(0.05, \mathcal{A}) \approx 0.0588$.
This illustrates how, even in a very simple set-up, and considering just two very similar testing procedures, the asymptotic size can be inflated by up to 18\% at the typical significance level of 0.05.
\end{Example}

\section{Proposed Framework}
\label{sec:method}

\subsection{Introducing the g-value}

To mitigate the impact of (subjective) analyst's choices on obtaining (objective) conclusions, we introduce a modified notion of p-value, which we call the {\em global p-value}, or \textit{g-value} in short. The g-value allows us to take into account a multiplicity of valid methods applicable to a given research problem, ensuring global consistency across them while preventing the risk of selective reporting. Specifically, our proposal selects the largest p-value among all the analyst's choices in $\mathcal{A}$ while adjusting the overall significance level to avoid excessive conservatism. In symbols, we define the g-value at significance level $\alpha$ as:
\begin{equation}
\label{eq:g-value}
   {G}_n (\mathcal{A}) \vcentcolon= \frac{\alpha}{\alpha^*(\alpha, \mathcal{A})} \max_{A \in \mathcal{A}} P_n(A) \, ,
\end{equation}
with
\begin{equation*}
\label{eq:correction}
    \alpha^*(\alpha, \mathcal{A}) \vcentcolon= \argzero_{\gamma \in [\alpha, 1]} \left[ \lim_{n \to \infty} \Pr \left\{ \max_{A \in \mathcal{A}} P_n(A) \leq \gamma  \big| H_0 \right\} - \alpha \right] \, .
\end{equation*}
The adjustment term $\frac{\alpha}{\alpha^*(\alpha, \mathcal{A})}$ provides an asymptotic correction to ensure that the procedure maintains an asymptotic size of $\alpha$. Similar ideas have been employed, for instance, in \cite{boulaguiem2024finite, boulaguiem2024multivariate} to adjust conservative procedures in the context of equivalence testing, a form of composite hypothesis testing. Moreover, analogous to classical testing procedures, the decision rule of the proposed approach is to not reject $H_0$ if $G_n(\mathcal{A}) > \alpha$ and to reject $H_0$ if $G_n(\mathcal{A}) \leq \alpha$. However, unlike the traditional p-value, the g-value depends on the significance level $\alpha$ through the adjustment term. 

\setcounter{Example}{0}
\begin{Example}[continued]
\label{ex:t-test:partII}
Revisiting our earlier example, we can express the test statistic of the one-sample t-test under $H_0$ as:
\begin{equation*}
    T_n^{(1)} = \frac{\sqrt{n} \bar{X}_n}{\hat{\sigma}_n} \, , 
    %\quad \text{with} \quad \bar{X}_n \vcentcolon= \frac{1}{n}\sum_{i=1}^n X_i \quad \text{and} \quad \hat{\sigma}_n^2 \vcentcolon= \frac{1}{n-1}\sum_{i=1}^n (X_i-\bar{X}_n)^2,  
\end{equation*}
where $\bar{X}_n \vcentcolon= \frac{1}{n}\sum_{i=1}^n X_i$ and $\hat{\sigma}_n^2 \vcentcolon= \frac{1}{n-1}\sum_{i=1}^n (X_i-\bar{X}_n)^2$,
and the test statistic of the Wilcoxon rank-sum test under $H_0$ as:
\begin{equation*}
    T_n^{(2)} = \sqrt{3}n^{-3/2}\sum_{i=1}^n R_i^+ \text{sign}(X_i) \, ,
\end{equation*}
where $R_i^+$ is the rank of $|X_i|$ and $\text{sign}(X_i)$ is 1 if $X_i\geq 0$ and $-1$ otherwise. Under regularity conditions, we can further write:
\begin{equation*}
    T_n^{(1)} = \frac{1}{\sqrt{n}}\sum_{i=1}^n \frac{X_i}{\sigma} + o_p(1) \quad \text{and} \quad
    T_n^{(2)} = \frac{\sqrt{3}}{\sqrt{n}}\sum_{i=1}^n  U_i \text{sign}(X_i) + o_p(1) \, ,
\end{equation*}
where $U_i\vcentcolon= G(|X_i|)$ and $G(\cdot)$ is the cumulative distribution function of $|X_i|$ \citep{lehmann1986testing}. %Then we 
We thus have:
\begin{equation}
\label{eqn:test_stat_sum}
    \begin{bmatrix}
    T_n^{(1)} \\
    T_n^{(2)}
    \end{bmatrix} = 
    \frac{1}{\sqrt{n}}\sum_{i=1}^n 
    \begin{bmatrix}
    X_i \sigma^{-1} \\
    \sqrt{3} U_i \text{sign}(X_i)
    \end{bmatrix} + o_p(1) \, .
\end{equation}
By the central limit theorem, we can show that: 
\begin{equation*}
    \frac{1}{\sqrt{n}}\sum_{i=1}^n 
    \begin{bmatrix}
    X_i \sigma^{-1} \\
    \sqrt{3} U_i \text{sign}(X_i)
    \end{bmatrix} \overset{d}{\to} \mathcal{N} \left(\0, \begin{bmatrix}
        1 & \sqrt{3/\pi} \\ \sqrt{3/\pi} & 1
    \end{bmatrix}\right),
\end{equation*}
and hence
\begin{equation*}
    \begin{bmatrix}
    T_n^{(1)} \\
    T_n^{(2)}
    \end{bmatrix} \overset{d}{\to} \mathcal{N} \left(\0, \begin{bmatrix}
        1 & \sqrt{3/\pi} \\ \sqrt{3/\pi} & 1
    \end{bmatrix}\right).
\end{equation*}
Next, based on this asymptotic joint distribution,
%of the two test statistics,
we can show that:
\begin{equation*}
\begin{aligned}
    \lim_{n \to \infty} \Pr \left\{ \max_{A \in \mathcal{A}} P_n(A) \leq \gamma  \big| H_0 \right\} &= \lim_{n \to \infty} \Pr \left[ \min_{A \in \mathcal{A}} \Phi\{T_n(A)\} \geq 1-\gamma  \big| H_0 \right] \\
    &= \lim_{n \to \infty} \Pr \left\{ \min_{A \in \mathcal{A}} T_n(A) \geq z_{1-\gamma}  \big| H_0 \right\} \\
    &= \gamma - \int_{z_{1-\gamma}}^\infty \varphi\left(y\right) \Phi\left(\frac{z_{1-\gamma} - \sqrt{3/\pi} y}{\sqrt{1 - 3/\pi}} \right) dy \, ,
\end{aligned}
\end{equation*} 
which allows us to express $\alpha^*(\alpha, \mathcal{A})$ as:
\begin{equation}
\label{eq:correction:closeform}
    \alpha^*(\alpha, \mathcal{A}) = \argzero_{\gamma \in [\alpha, 1]} \left\{ \gamma - \int_{z_{1-\gamma}}^\infty \varphi\left(y\right) \Phi\left(\frac{z_{1-\gamma} - \sqrt{3/\pi} y}{\sqrt{1 - 3/\pi}} \right) dy - \alpha \right\} \, .
\end{equation} 
For instance, for $\alpha=0.05$, we obtain $\alpha^*(0.05, \mathcal{A}) \approx 0.0601$. Note that our approach here can be rendered equivalently as: 
\begin{itemize}
    \item[(i)] taking the largest p-value and comparing it to $6.01\%$; or
    \item[(ii)] taking the largest p-value, correcting it by $0.05/\alpha^*(0.05, \mathcal{A}) \approx 0.832$, and comparing it to $5\%$.
\end{itemize}
\end{Example}

\subsection{A numerical illustration}
\label{sec:toy_simu}

To illustrate the empirical performance of our proposal, we generate a random sample $\{X_i\}_{i=1,\ldots,20}$ of size $n=20$ with $X_i - \mu \overset{iid}{\sim} t_{15}$, where $t_{15}$ denotes a Student's t-distribution with 15 degrees of freedom, and $\mu \in [0,1]$. We aim to test the hypotheses formulated in \eqref{eq:hypotheses}. Hence, when $\mu=0$, we can evaluate the type I error rates of different testing procedures, and when $\mu>0$ we can evaluate their statistical powers. 

We consider again an $\mathcal{A}$ comprising (i) the {\em one-sample t-test}, and (ii) the {\em Wilcoxon rank-sum test}. These two choices are then combined in three possible ways, namely: (iii) with a {\em liberal approach}, 
where one takes the smallest of the two p-values from (i) and (ii); (iv) with a {\em conservative approach}, 
where one takes the largest of the two p-values from (i) and (ii); and (v) with the {\em g-value approach} in \eqref{eq:g-value} -- that is, by correcting the largest of the two p-values from (i) and (ii) with the adjustment term $\alpha^*(\alpha, \mathcal{A})$ in \eqref{eq:correction:closeform}. 

\begin{figure}[!h]
    \centering
    \includegraphics[width=\textwidth]{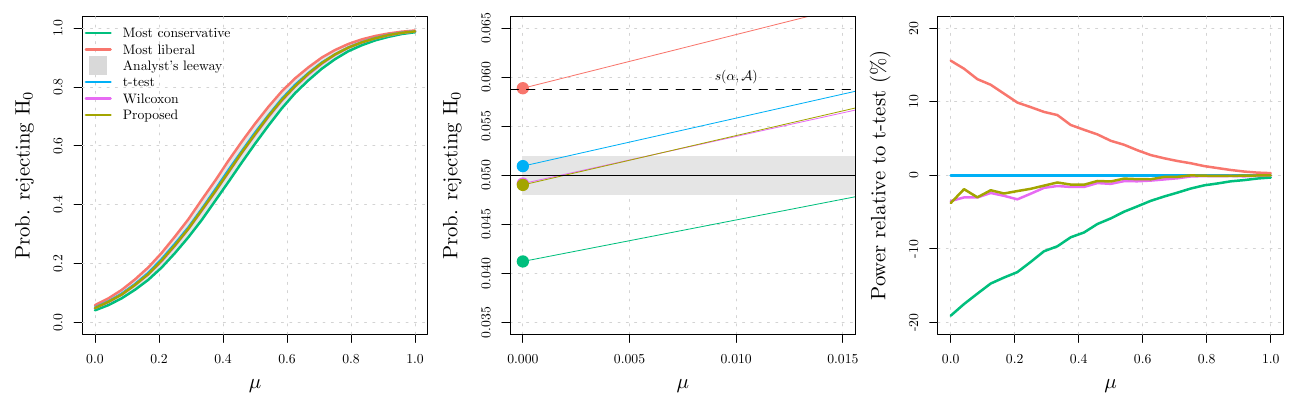}
    \caption{Performance comparison on simulated data for the testing procedures described in Section~\ref{sec:toy_simu}}
    \label{fig:toy_simu}
\end{figure}

The performance of (i)-(v) in $10^5$ Monte-Carlo replications is summarized in Figure~\ref{fig:toy_simu}. The first panel shows the probability of rejecting $H_0$ as a function of $\mu$. A steeper curve indicates greater power. As expected, the liberal approach exhibits the highest power and the conservative approach the lowest; the space between them illustrates the leeway available to the analyst -- that is, the ``power spread''  created by one's choice. The second panel shows a close-up around $\mu=0$. At a significance level of $\alpha=5\%$, the liberal approach exhibits a type~I error rate inflated to $s(\alpha, \mathcal{A}) \approx 0.0588$ (see calculation in Example~\ref{ex:t-test}) and the conservative approach exhibits a type I error rate well below the nominal level. In contrast, the two original test procedures and the g-value approach maintain type I error rates close to 0.05, within an acceptable range of computational error. This suggests that our proposal properly controls the type I error rate, even with small sample sizes. The third panel illustrates the relative power improvement compared to the one-sample t-test, measured by the relative difference in the probability of rejecting the null when the alternative is true. Notably, the power of the g-value approach closely matches that of both the t-test and the Wilcoxon test. 

In summary, these results suggest that integrating multiple statistical strategies into a unified framework is feasible. While the liberal approach increases the risk of false positives and the conservative approach increases the risk of false negatives, our g-value proposal effectively balances and controls both risks, maintaining valid inference with only a minor loss in efficiency.

\section{Practical Considerations}
\label{sec:practical}

\subsection{Construction of $\mathcal{A}$}
\label{sec:constructA}

The considered set of analyst's choices for a given research problem, $\mathcal{A}$, plays a crucial role in determining the applicability and effectiveness of our framework.
%is to identify the set $\mathcal{A}$ of potential analysis strategies given a research problem, which
In principle, $\mathcal{A}$ may comprise alternatives spanning the entire process -- from data collection, to data cleaning and preprocessing, to choices of models and/or algorithms. Not accounting for such alternatives can increase the risk of biased and non-replicable conclusions.

Although there is seldom a single ``correct'' way to perform an analysis,
%. As George E.~P.~Box famously observed, ``\textit{all models are wrong, but some are useful}''
%.'' 
%Nevertheless, not taking alternative strategies into account can cause an even bigger issue, such as increasing the risk of biased conclusions. Although it is generally 
it may be simply infeasible to enumerate all possible meaningful and reasonable strategies. 
%not accounting for alternatives can increase the risk of biased conclusions. 
Following George E.~P.~Box's famous observation that ``\textit{all models are wrong, but some are useful}'', we believe one should strive for a middle ground, with the main objective to foster transparency. By requiring that they explicitly outline the set $\mathcal{A}$, we compel researchers to thoroughly consider the full spectrum of legitimate analysis pathways when addressing their research question, thereby helping to prevent selective reporting or cherry-picking of favorable results. 

Below we outline several practical perspectives that may be considered when constructing $\mathcal{A}$. 

\begin{itemize}
    \item \textbf{Cleaning and preprocessing of the data:} When given a dataset, researchers typically begin their work with exploratory steps to understand its basic characteristics and address any cleaning and preprocessing needs. Among these is typically the handling of missing values \citep{allison2009missing} -- which can be handled by removing observational units and/or variables, or through one of many available  imputation techniques \citep{miao2022experimental}. Additionally, researchers often handle outliers, for which a broad variety of techniques exist as well \citep{wang2019progress,boukerche2020outlier}.
    \item \textbf{Variable transformations:}
    In many studies, researchers also need to implement transformations on one or more variables. For example, when analyzing volume measurements, such as tumor size in oncology, a cube root transformation is commonly used to stabilize variance and facilitate interpretation. In other situations, transformations may be required for statistical reasons, such as normalizing data to ensure that model errors are identically distributed. These choices are also critical and must be carefully considered as part of the analysis pipeline.
    \item \textbf{Variable selection:}
    Many studies also require variable selection, e.g., the selection of predictors to be employed in regression-type models or predictive algorithms. This can have a substantial impact on the final conclusions. Consequently, which techniques are utilized, e.g., a LASSO \citep{tibshirani1996regression}, adaptive LASSO \citep{zou2006adaptive} or SCAD \citep{fan2001variable}, among many others, and how such techniques are tuned become critical considerations.
    \item \textbf{Choice of statistical models and testing procedures:}
    Different statistical tools can be equally valid for addressing a specific question through a given dataset. For example, when testing for a population mean, both the t-test and the Wilcoxon rank-sum test are legitimate options. In more complex regression settings, analysts must carefully choose which models to consider and evaluate. It is essential to support these choices with model validity testing procedures, such as the Sargan–Hansen J-test commonly used in the generalized method of moments literature \citep{hansen1982large}, to help identify and exclude mis-specified models.
\end{itemize}

Depending on the specific study, the cardinality of a reasonable $\mathcal{A}$ can be quite substantial -- due to the wide range of options available. This increases the risk that the computation of the g-value, which is based on the maximum of the p-values across {\em all} $A$'s in $\mathcal{A}$, becomes overly sensitive to the inclusion of a few poorly fitting analyst's choices. Such sensitivity can undermine the accuracy and reliability of the inference based on the g-value. To mitigate such sensitivity and increase the robustness of our approach, we propose to compute an alternative g-value by replacing the maximum with a specified quantile of the p-values, i.e.,
\begin{equation*}
   {G}_n (\mathcal{A}; \gamma) \vcentcolon= \frac{\alpha}{\alpha^*(\alpha, \mathcal{A})} \text{Quantile}\left\{\gamma; P_n(A), A\in\mathcal{A} \right\} \quad \text{with} \quad \gamma\in(0,1] \, .
\end{equation*}
This reduces the influence of extreme or outlying analyst's choices, thereby providing a more robust calculation of the g-value based on the constructed $\mathcal{A}$, while preserving interpretability and statistical validity.

\subsection{Computation of g-value}
\label{sec:computation}

A key aspect of computing the g-value is obtaining the asymptotic joint distribution of the p-values, or that of the test statistics, associated with each $A$ in $\mathcal{A}$ under the null hypothesis. This is what allows us to derive $\lim_{n \to \infty} \Pr \left\{ \max_{A \in \mathcal{A}} P_n(A) \leq \gamma  \big| H_0 \right\}$, which is needed for the computation of the g-value. In our Example~\ref{ex:t-test:partII}, we expressed both test statistics as sample averages, as shown in \eqref{eqn:test_stat_sum}, which allowed us to apply the central limit theorem to derive their asymptotic joint distribution under the null hypothesis. The same principle can be extended to regression settings. Specifically, suppose that we observe an independent sample $\{Y_i,\x_i\}_{i=1,\ldots,n}$, based on which we obtain an estimator $\hat{\bb}$ for the true regression coefficient vector $\bb$ of dimension $p$. Our interest lies in testing $H_0: \beta_j=0$ vs.~$H_1:\beta_j>0$ for some $j\in\{1,\ldots,p\}$. Moreover, suppose that $\hat{\bb}$ satisfies:
\begin{equation*}
    \sqrt{n}\left(\hat{\beta}_j-\beta_j
    %\beta_{0,j}
    \right) = \sum_{i=1}^n \phi^{(A)}_j(Y_i, \x_i, 
    %\bb_0
    \bb) + o_p(1) \quad \text{with} \quad j=1,\ldots,p \, ,
\end{equation*}
where $\phi^{(A)}_j(\cdot)$ is some scalar function computed on the data and the true parameter, and is based on the $A\in\mathcal{A}$ which is being used for inference. This is commonly satisfied, for example, when $\hat{\bb}$ is the MLE. We can then use standard techniques (e.g., the central limit theorem) to obtain the asymptotic joint distribution of $\{\sum_{i=1}^n\phi^{(A)}_j(Y_i, \x_i, \bb)\}_{A\in\mathcal{A}}$ under the null hypothesis, which allows us in turn to obtain the one of the p-values. 

For more complex scenarios, deriving a closed-form expression for the asymptotic joint distribution of the p-values may be intractable. In such cases, a solution is to shift from analytical derivations to numerical approximations using resampling methods. The main difficulty then lies in ensuring that the resampled datasets satisfy the null hypothesis for all analyst's choices comprised in $\mathcal{A}$. To overcome this, and still considering independent observations, we propose an algorithm that combines nonparametric bootstrap with data permutation. Specifically, we first resample the data with replacement, and then permute the values of the variable of interest. The permutation step disrupts the dependence between the response variable and the regressors, thereby ensuring that the null hypothesis holds for all resampled datasets across all $A$'s in $\mathcal{A}$.

We note that the case of dependent observations (e.g., time series) introduces additional challenges for deriving the asymptotic joint distribution of the p-values, both analytically and numerically. Addressing dependent data is beyond the scope of this paper and is left for future research.

\section{Conclusions}
\label{sec:conclusion}

The replicability crisis has impacted a wide range of scientific disciplines, largely due to questionable research practices such as selective reporting and p-hacking. These issues undermine both the advancement of science and public trust in scientific findings. Various approaches have been proposed to detect and address the replicability crisis, among which multi-design studies have attracted significant attention for their potential to incorporate diverse analytical perspectives. Nevertheless, the existing literature lacks unified frameworks for integrating the numerous resulting outcomes, particularly when these outcomes are in the form of p-values.

To address this gap, we introduced the g-value, a novel measure that leverages the complete spectrum of p-values derived from multiple legitimate analyst's choices to answer a single research question. Our findings demonstrate that the g-value controls the type I error rate at the nominal level while maintaining strong statistical power. This indicates that it is feasible to robustly integrate multiple analysis strategies in a unified manner. We showcased the practical usefulness of the g-value through a simple and classical hypothesis testing problem for the population mean. Our approach reduces the risk of selective reporting, contributing to more transparent, trustworthy, and replicable scientific research.

\newpage
\normalem % for italic journal name in reference
\bibliographystyle{apalike}
\bibliography{bib}
\end{document}